\documentclass[final]{IEEEtran}
\usepackage{amsmath}
\usepackage{amsthm}
\usepackage{multirow}
\usepackage{color,cite}
\usepackage{subfigure}
\usepackage{mathrsfs}
\usepackage{amssymb}
\usepackage{bm}
\usepackage{graphicx}
\usepackage{amsfonts}
\usepackage{algorithm}
\usepackage{algorithmic}
\usepackage{makecell}
\usepackage{etoolbox}

\newtheorem{theorem}{Theorem}
\newtheorem{lemma}{Lemma}

\allowdisplaybreaks[4]

\input epsf

\begin{document}

\title{{Doppler-Resilient Design of CAZAC Sequences for \\ mmWave/THz Sensing Applications}}

\author{Fan Zhang, Tianqi Mao,~\IEEEmembership{Member,~IEEE}, Zhaocheng Wang,~\IEEEmembership{Fellow,~IEEE}

\thanks{This work was supported by the National Key R\&D Program of China under Grant 2018YFB1801102.
\emph{(Corresponding author: Zhaocheng Wang, Tianqi Mao)}}
\thanks{F. Zhang and Z. Wang are with Department of Electronic Engineering, Tsinghua University, Beijing 100084, China. Z. Wang is also with the Shenzhen International Graduate School, Tsinghua University, Shenzhen 518055, China. (E-mails: zf22@mails.tsinghua.edu.cn, zcwang@tsinghua.edu.cn).}
\thanks{T. Mao is with the School of Electronic and Information Engineering, Beihang University, Beijing 100191, China (e-mail:maotq@buaa.edu.cn).}
 \vspace{-10mm}
 }

{}

\maketitle

\begin{abstract}
%
Ultra-high-resolution target sensing has emerged as a key enabler for various cutting-edge applications, which can be realized by utilizing the millimeter wave/terahertz frequencies.
%
However, the extremely high operating frequency inevitably leads to significant Doppler shift effects, especially for high-mobility applications, causing the degradation of sensing performance with high false alarm rate.
%
 To this end, this paper proposes a parameter design methodology of the well-known constant amplitude zero auto correlation (CAZAC) sequences, which aims at enhancing their resilience to Doppler shifts. 
 Specifically, we suppress the sidelobes incurred by Doppler shifts for the peak-to-sidelobe ratio (PSLR) improvement within the range of interest (RoI) of the radar range profile\footnote{The radar range profile is referred to as the cross-correlation between the transmitted sequence and the received echos.}.
 %
%
The Zadoff-Chu (ZC) sequence, as a representative member in the CAZAC family, is firstly considered.
The impacts of its root index on range sidelobes are investigated based on number theory. For an arbitrary-length ZC sequence, a feasible range of the root index is derived to satisfy the requirement of PSLR within the scope of RoI. 
%
Furthermore, these design guidelines are extended to a general form of CAZAC sequences, where a low-complexity heuristic algorithm is developed for PSLR improvement. 
Simulation results demonstrate that under severe Doppler shifts, our proposed methodology could enhance the sensing performance by lowering the false alarm rate while maintaining the same 
detection rate, compared with its classical counterpart.
\end{abstract}


\IEEEpeerreviewmaketitle

 \vspace{-4mm}
\section{Introduction}
Ultra-high-resolution target sensing has drawn much attention in diverse 
technologies including autonomous driving, virtual
reality, and Internet of Things (IoT)\cite{ref1,ref2}.
%
The stringent resolution requirement prompts the exploration of the millimeter wave/terahertz (mmWave/THz) bands, which could improve the resolution of the target range and velocity estimation due to their wide operation bandwidth and high carrier frequency\cite{ref3}.
%
On the other hand, with the rapid development of digital signal processing technology, the structures of communication and sensing transceivers have become increasingly similar. Therefore, the mmWave/THz radar sensing can be integrated into 5G/6G mmWave/THz communication systems with low overhead, which forms the integrated sensing and communication (ISAC) systems\cite{ref4}.

Constant amplitude zero auto correlation (CAZAC) sequences, represented by Zadoff-Chu (ZC) Sequence\cite{ref6}, Frank sequence\cite{ref7}, and chirp sequence\cite{ref8}, are considered to be promising candidates for accurate radar sensing, thanks to their advantages of low peak-to-average power ratio (PAPR) and perfect auto-correlation (AC) property. 
 {However, in mmWave/THz sensing applications, the extremely high carrier frequency and the long sensing sequence which is required to combat the high pathloss, lead to significant Doppler shift effects.} 
The strong Doppler shifts can destroy the perfect AC property of CAZAC sequences, which induces non-negligible sidelobes on the radar range profile, leading to degraded the sensing performances with high false alarm rate.

To tackle this issue, there exists some literature enhancing the Doppler resilience of CAZAC sequences{\cite{ref14,ref11,add1}}. 
 {A Doppler-resilient code for MIMO radar was designed  in \cite{add1} at the expense of redundancy in the slow time domain, which makes range sidelobes vanish at modest
Doppler shifts.}
The design of Doppler-resilient Golay pairs was investigated 
in \cite{ref14}. 
Although it could suppress the range sidelobes incurred by Doppler shifts, it was hard to extend to other CAZAC sequences.
Besides, the differential ZC (DZC) sequence proposed in \cite{ref11} preserved the quasi-perfect AC property in the presence of severe Doppler shifts, which, unfortunately, exhibited poor sensing
performance at low signal-to-noise-ratio (SNR).

Against this background, this paper proposes a parameter design methodology to derive Doppler-resilient CAZAC sequences for high-mobility applications, e.g., autonomous driving\footnote{In addition to autonomous driving, which is taken as an example in this paper, the proposed design can be employed in various sensing applications.}.
%
Considering that high path loss limits the coverage range of mmWave/THz bands, their sensing range is typically much smaller than the length of the whole radar range profile\cite{range,range1}.
%
Accordingly, instead of considering the whole profile, our proposed methodology aims at maintaining high PSLR within the range of interest (RoI), which is relevant to the sensing range of practical applications.
Specifically, the root index of ZC sequences is firstly investigated based on number theory.
For a ZC sequence of arbitrary length, a feasible range of the root index is derived to satisfy the required PSLR threshold within 
RoI.
By extending the design guidelines to a general form of CAZAC sequences, a low-complexity algorithm is proposed to optimize the sequence parameters for improving the PSLR within 
RoI.
Simulation results demonstrate that under severe Doppler shifts, our proposed methodology could achieve a lower false alarm rate than its state-of-art counterpart with the same level of detection rate.

 \vspace{-4mm}
\section{System Model}
\begin{figure}[tp!] 
\vspace{-5mm}
	\begin{center}
		\includegraphics[width=0.45\textwidth]{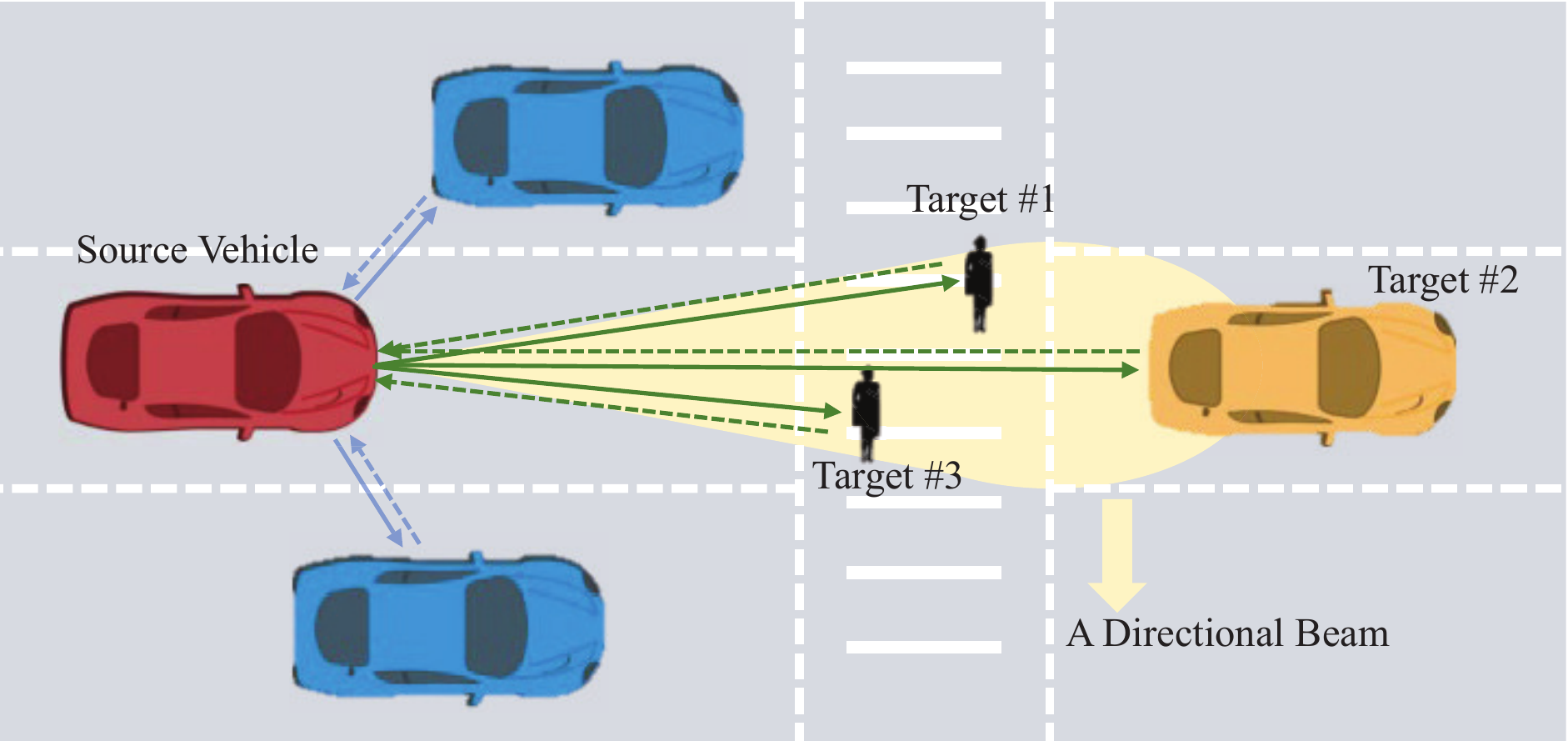}
	\end{center}
	\caption{The multi-target sensing scenario in autonomous driving.}
	\label{fig2}
	\vspace{-5mm}
\end{figure}



Figure \ref{fig2} illustrates a multi-target sensing scenario in autonomous driving applications, where a source vehicle (in red) keeps emitting the { $N$-length} sensing sequence $\mathbf{s}\!=\!\! [s[0],s[1],\cdots,s[N-1]]$ to different directions, and captures the echoes for radar detection. 
To combat the severe path loss at mmWave/THz frequencies, a large-scale antenna array with highly directional beams is often employed, which is equivalent to one directional antenna (e.g., horn antenna). Hence, we consider the single-input single-output (SISO) case for simplicity.

Assuming that there are a total of $L$ targets, where the relative distance and velocity to the source vehicle of the $l$-th target $(l=1,2,\cdots,L)$ are denoted as $d_l$ and $u_l$, respectively.
Then, the round-trip delay of the echo from the $l$-th target can be expressed as
\begin{equation}
	\begin{aligned}
		t_l= \frac{2d_l}{c},
	\end{aligned}
	\label{eq2}
\end{equation}
where $c$ denotes the speed of light.
Besides, the normalized Doppler shift of the $l$-th target can be formulated as
\begin{equation}
	\begin{aligned}
		{v}_l=\frac{2u_lf_c}{c}{T_s},
	\end{aligned}
	\label{eq2}
\end{equation}
where $f_c$ and $T_s$ denote the carrier frequency and the sampling period, respectively. Since most roads have the speed limits for vehicles, we assume that the absolute value of $u_l$ is no more than the maximum relative velocity $\bar{u}$ ($|u_l| \leqslant \bar{u}$) and hence $|v_l| \leqslant \bar{v}$ ($\bar{v} =\frac{2\bar{u}f_c}{c}{T_s}$).
The reflected signal received by the source vehicle in a discrete form, denoted as $\mathbf{y}=\left [y[0],y[1],\cdots,y[N-1]\right ]$, can be expressed as
\begin{equation}
	\begin{aligned}
		y[n] =\sum_{l=1}^{L}h_l s[n-\tau_l]e^{\textsf{j}(2\pi n{v}_l)}+ {w}[n],
	\end{aligned}
	\label{eq1}
\end{equation}
where $h_{l}$ is the round-trip path gain of the $l$-th target and ${\tau}_l=\lfloor\frac{t_l}{T_s}\rfloor$ is the integer round-trip delay. {In the discrete time domain, since integer time delays are much easier to depict than  fractional delays, for simplicity, fractional time delays are rounded to the nearest integers, which is commonly used in studies on radar sensing systems \cite{ref15,ref12}.}
Besides, ${w}[n]$ represents the noise component caused by thermal noise and the clutters from other directions, following complex Gaussian distribution, i.e., $w[n] \sim \mathcal{CN}(0,\sigma^2)$.  

By repetitively transmitting $\mathbf{s}$ for $K$ times, the radar range profile for the $k$-th transmission ($k=1,2,\cdots,K$) can be calculated in a circular correlation form as
\begin{align}
	  r^{(k)}[n]  & =  \sum_{i=0}^{N-1} y_k[i]s^*[\langle i-n \rangle_{N}],
\end{align}
where $\mathbf{y}_k$ denotes the received echo signal for the $k$-th transmission. Besides, $(\cdot)^*$ and $\langle \cdot \rangle_{N}$ denote the conjugate operator and the $N$-modulo operator, respectively. For brevity, the notation  $\langle \cdot \rangle_{N}$ is omitted for sequence indices in the rest of the paper.
Afterwards, the range-Doppler-matrix (RDM) $E(n,q)$ can be obtained by performing DFT on the correlation values, {which is can be expressed as
\begin{align}
	  E(n,q) =  \sum_{k=0}^{K-1} r^{(k)}[n] e^{\textsf{j}\frac{-2\pi kq}{K_0}} .
\end{align}
$n$ and $q$ are the indices in the time and Doppler domains, respectively, with $n=0,1,\cdots, N-1$ and $q=0,1,\cdots, K_0-1$. Besides, $K_0$ is the FFT size with $K_0 = \omega K$, where $\omega$ is a positive integer number. 
$E(n,q)$ is then employed to detect targets via hypothesis tests, which can be written as 
\begin{align}
	 \text{H}_1:\frac{||E(n,q)||^2}{\theta(n,q)} >\Gamma,\quad \text{H}_0:\frac{||E(n,q)||^2}{\theta(n,q)} <\Gamma.
\end{align}
 $\text{H}_1$ is the hypothesis that the point $(n,q)$ corresponds to a true target and $\text{H}_0$ is the hypothesis that the point $(n,q)$ is a noise or clutter point. $\theta(n,q)$ is the average noise power estimation at the point $(n,q)$, which is calculated by averaging the value of $||E(n,q)||^2$ in the cell $[0,N-1] \times [0, K_0-1]$ excluding $(n,q)$ \cite{ref15}. 
 In the simulations, we adjust the threshold $\Gamma$ to obtain different false alarm rates and detection rates for plotting the receiver operating characteristic (ROC) curve.

For a given point $(n_0,q_0)$, if the hypothesis $\text{H}_1$ holds, the integer round-trip delay $\tau$ and normalized Doppler shift $v$ can be expressed as $\tau = n_0$ and $  v = \frac{q_0}{N K_{0}}$, respectively. Since the sampling rate is limited, i.e., both $n_0$ and $q_0$ must be integers, the permissible error range for the relative distance and velocity are $[0,\frac{cT_s}{4})$ and $[0,\frac{c}{4NK_0T_sf_c})$, respectively \cite{ref12}.} 


\section{Proposed Parameter Design Methodology}
%
\begin{table*}[t]
	\vspace*{-5mm}
	\renewcommand{\arraystretch}{1.1}
	\caption{The Corresponding Relationship Between $|n-\tau|$ and $| \langle p (n -\tau) \rangle _{N }|$.}
	\begin{center}
			\begin{tabular}{|c | c | c | c | c | c | c | c | c | c | c |c|}
					\hline
					$ |n-\tau |$   & $0$ & $ 1$ & $\cdots$ & $  A-1$  & $ A$ & $  A+1$&$  A+2$&$\cdots$& $  2A-1$ & $ 2A$& $\cdots$\\\hline
					
					$\left|\left \langle  p(n-\tau)\right \rangle_{N }\right|$  & 0 & $p$ &  $\cdots$ &$(A-1)p$ &$Ap$& $(\! A \! -\! 1\!)p+2B+1$ & $(\! A \!-\! 2)p+2B+1$ & $\cdots$&$p+2B+1$ &$2B+1$& $\cdots$ \\\hline
 		 
				\end{tabular} 
		\end{center}
	\vspace*{-5mm}
	\label{t1}
\end{table*}

\subsection{Parameter Design for ZC Sequences}
Without loss of generality, consider that the odd-length ZC sequence $\mathbf{s}$ is utilized as the radar sensing sequence, which can be expressed as
\begin{align}
	s[n] = \exp(-\textsf{j} \frac{pn(n+1)\pi}{N}),
\end{align}
where $p$ and $N$ denote the root index and the sequence length (odd), respectively. 
{The root index $p$ satisfies $0<p<N $ and $\gcd(p,N)=1$, where $\gcd(\cdot)$ is the greatest common divider calculator.} 
The perfect AC property of ZC sequences can be written as 
\begin{align}
	  r_s[n]   \! =  \sum_{i=0}^{N -1} s[i-\tau] s^*[i-n]   = 
\begin{split}\textit{}
	\begin{array}{rcl}
		N,\quad
		\text{$n= \tau$};\\
		0,\quad
		\text{$n \neq \tau$},
	\end{array}
\end{split}
\end{align}
where $n$ denotes the index of correlation results ($n = 0,1,\cdots,N-1$) and $\tau$ denotes the integer round-trip delay.
Accordingly, the radar range profile in multi-target sensing only exhibits several main peaks corresponding to the targets but no dominant sidelobes when Doppler shifts are marginal.
\begin{figure}[b]
\vspace{-6mm}
	\center
	\includegraphics[width=0.7\linewidth, keepaspectratio]{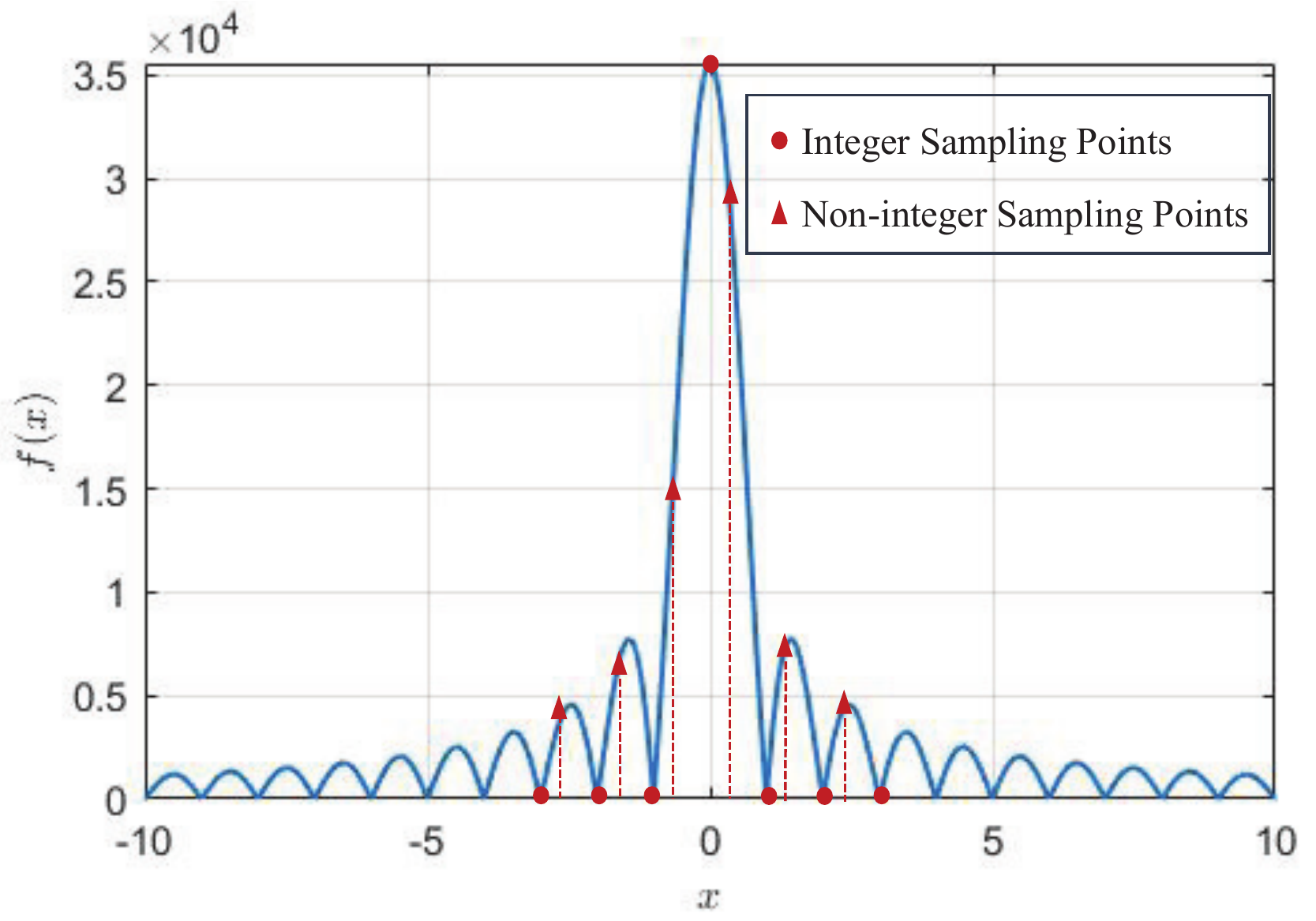}	
	\caption{The graph of $f(x)$ with respect to $x$, where $N=35537$.}
 
 \label{sin_sin}
	\vspace{-5mm}
\end{figure}

However, the Doppler shift effects tend to be strong under high-mobility scenarios at mmWave/THz frequencies, which impairs the perfect correlation property of the ZC sequence, inducing high sidelobes on the radar range profile.
Specifically, for each target, the magnitude of cross-correlation between the transmitted ZC sequence and the echo signal with the normalized Doppler shift $v$ and the integer round-trip delay $\tau$ can be derived as 
\begin{align}
	 \left \| r_{s}[n] \right \|& =\left \| \sum_{i=0}^{N-1} s[i-\tau]s^*[i-n]e^{\textsf{j}2\pi i {v}} \right \| \nonumber  \\
	& = \left  \| \frac{\sin\left(\pi \langle p (n -\tau)   +{v}N \rangle _{N } \right) }{\sin\left(\frac{\pi}{N}\langle p (n -\tau)   +{v}N \rangle _{N }  \right)} \right \|,  \label{auto-zc}  
\end{align}
where $\|\cdot\|$ denotes modulus of complex number.
Note that the result derived by $\langle \cdot \rangle_{N}$ here could be a fraction number.
Without loss of generality, the principal value range of  $\langle \cdot \rangle_{N}$ is constrained in $ [- \frac{N-1}{2} , \frac{N-1}{2}] $.  

According to (\ref{auto-zc}), 
$\left \| r_{s}[n]\right \|$ 
can be regarded as the function of
$ f(x) = \left\| \frac{\sin (\pi x)}{\sin(\frac{\pi}{N } x)}\right \|$ sampled on $x = \langle p (n -\tau) +{v}N  \rangle _{N }$.
When the Doppler shift is negligible, i.e., $v \approx 0$, $f(x)$ is sampled on the integer points. Hence, $\left \| r_{s}[n]\right \|$  equals $N$ as $x$ approaches zero (corresponding to $n = \tau$), and $0$ for all the other cases. Nevertheless, when 
$v\neq 0$, the sampling points will deviate from integer points, resulting in the appearance of high sidelobes, as shown in Fig.~\ref{sin_sin}.

Meanwhile, $ \langle p (n -\tau) \rangle _{N}$ for $n\!\!=\!\!0,1,\cdots, N\!\!-\!\!1$ constitutes the complete residue system modulo $N$ denoted as $\mathcal{Q}=\{-\frac{N-1}{2},-\frac{N-1}{2}+1,... ,\frac{N-1}{2}\}$. Each element of $\mathcal{Q}$ corresponds to a unique sidelobe on the range profile, whilst the zero value corresponds to the main peak. Inspired by these properties, the positions of sidelobes on the range profile can be modified by adjusting the root index $p$, which actually changes the permutation of the sequence $ \langle p (n -\tau) \rangle _{N}$ for $n\!=\!0,1,\cdots, N\!-\!1$. 
The permutation of the sequence $| \langle p (n -\tau) \rangle _{N}|$ with respect to $|n-\tau|$ is investigated in Table \ref{t1}, where we denote $\lfloor \frac{N-1}{2p} \rfloor$ as $A$ and $\frac{N-1}{2} - Ap$ as $B$ for brevity.
For simplicity, we assume that $N$ is a prime, but all the analysis can be naturally extended to any non-prime values of $N$ by excluding the values of $p$ with $\gcd(p,N) \neq 1$ from the candidates. 

Since the sensing range in practical applications is typically much smaller than the length of the radar range profile, it is not necessary to consider the whole profile. Therefore, 
based on the sensing range $D_{\rm{r}}$ for different sensing applications, we define RoI as $0 < n < \frac{2D_{\rm{r}}}{cT_s}$, where $\frac{2D_{\rm{r}}}{cT_s}$ is the maximum round-trip delay within $D_{\rm{r}}$.
Our proposed Doppler-resilient design aims at improving the PSLR within the scope of RoI, where the PSLR can be formulated as
\begin{align}
\text{PSLR} \overset{\text{def}}{=}  \frac{\left \| r_{s}[\tau]\right \| }{\max\limits_{ n \neq \tau} \left \| r_{s}[n]\right \|} .
\end{align}
Specifically, we design the root index $p$ to remove the high sidelobes out of RoI so that PSLR can be higher than the required PSLR threshold $P_r$, which can be formulated as
\begin{align}
	\text{find} &\quad p   \nonumber \\
	\text{s.t.} \quad	\!\frac{\left \| r_{s}[\tau]\right \| }{\max\limits_{ n \neq \tau} \left \| r_{s}[n]\right \|}\!\! \geqslant & P_r, \text{ for $0 < n < \frac{2D_{\rm{r}}}{cT_s} $}. \label{problem1}
\end{align}
{
It should be noted that by moving the high sidelobes away from the main peak, our proposed root index design can also narrow the width of the main lobe, thus maintaining high detection accuracy even in the presence of sampling errors.}

Due to the complex expression of $\left \| r_{s}[n] \right \|$, it is difficult to  solve the problem (\ref{problem1}) directly.
Fortunately, a sufficient condition for the solutions of (\ref{problem1}) can be derived based on Theorem \ref{the1}.
\begin{theorem}
When a ZC sequence of length $N$ with the root index $p$ is employed for radar sensing, the PSLR satisfies 
\begin{align}
 \frac{\left \| r_{s}[\tau]\right \| }{\max\limits_{ n \neq \tau} \left \| r_{s}[n]\right \|} =
  \| \frac{\sin\left(\frac{\pi}{N }\left( p  +\bar{v}N \right) \right)}{\sin\left(\pi \bar{v} \right) } \|, \quad \text{for $n \in \mathcal{S}_p$}, \label{lobe}
\end{align}
where $\mathcal{S}_p\! = \! \{n \in \mathbb{Z}| 0\!<\!| n\!-\!\tau  |\!<\!2 \lfloor \frac{N-1}{2p} \rfloor \}$ and $\bar{v}$ is the maximum normalized Doppler shift with $\bar{v}N<1$\footnote{Due to the short sampling period in mmWave/THz systems, $v$ is around $10^{-6} \sim 10^{-5}$. Besides, instead of directly using a long sequence, this paper considers periodical transmission of short sensing sequences. Therefore, the assumption $vN<1$ fits well with many practical applications \cite{ref12},\cite{add2}.}
\label{the1}
\end{theorem}
Theorem \ref{the1} can be proved based on Table \ref{t1} and the nature of $\left \| r_{s}[n]\right \|$. Firstly, it can be seen from Table \ref{t1} that $| \langle p (n -\tau) \rangle _{N}| \geqslant p$ when $n \in \mathcal{S}_p$. 
By assuming that $vN<1$, $\left \| r_{s}[n]\right \|$  decreases as the $| \langle p (n -\tau) \rangle _{N}|$ becomes larger. 
It is evident that the highest sidelobe within $\mathcal{S}_p$ corresponds to $| \langle p (n -\tau) \rangle _{N}|=p$.
By substituting $| \langle p (n -\tau) \rangle _{N}|= p$  into (\ref{auto-zc}), the PSLR within $\mathcal{S}_p$ can be calculated as (\ref{lobe}).

According to Theorem \ref{the1}, a sufficient condition for the feasible solutions of (\ref{problem1}) can be derived as below. If there exists $p$ such that $\mathcal{S}_p$ contains RoI and the PSLR within $\mathcal{S}_p$ is greater than the threshold $P_r$, the value of $p$ could be a solution to (\ref{problem1}), which can be expressed as
\begin{align}
\text{find}&\quad p   \nonumber \\
\text{s.t.} \quad (0,\frac{2D_{\rm{r}}}{cT_s}) \subseteq \mathcal{S}_p \!\quad\! & \land \!\quad\! \| \frac{\sin\left(\frac{\pi}{N}\left( p  +\bar{v}N \right) \right)}{\sin\left(\pi \bar{v} \right) } \|   \geqslant  P_r. \label{sufficient} 
\end{align}
By solving (\ref{sufficient}), a feasible range of $p$ in (\ref{problem1}) can be derived as
\begin{align}
	\!\frac{N}{\pi}(\arcsin(P_{r} \sin(\pi \bar{v})) \! - \! \pi \bar{v}) \! \leqslant \!
	p \! \leqslant \! \frac{(N \! - \! 1 \! - \! 2B)cT_s}{2D_{r}}. \label{solution}
\end{align}
{In the case of non-prime $N$, only the values in (\ref{solution}) satisfying $\gcd(p,N)= 1$ can be selected as feasible solutions.
}
 

Additionally, since the PSLR within $\mathcal{S}_p$ is a {monotonically increasing}  function with respect to $p$ when $p<\lfloor N/2 \rfloor$, 
{among all the feasible solutions, the largest $p$ makes PSLR reach the maximal. Specifically, for any prime $N$, the largest $p$ within the feasible solutions can be expressed as } 
\begin{align}
\label{solution1}
p=\lfloor \frac{(N -1-2B)cT_s}{2D_{r}} \rfloor .
\end{align} 
{For non-prime $N$, the largest $p$ within the feasible solutions should be the value that is closest to and less than (\ref{solution1}).}

{ Note that the Doppler shift effects also cause ambiguity in the joint estimation of range and velocity. To avoid ambiguity, the search range of Doppler shifts is limited based on the prior information of Doppler shifts, which will be further clarified in Section~\uppercase\expandafter{\romannumeral4}.}

\vspace{-2mm}
\subsection{Extension to General CAZAC Sequences}
\label{extension}
Inspired by the root index design on ZC sequences, we further extend the proposed design methodology  to the general form of CAZAC sequences. 
\subsubsection{A Unified Construction of CAZAC Sequences}
\label{PRUS}
All the members of the CAZAC sequence family can be constructed by a unified method shown as Lemma \ref{lem-1}.
\begin{lemma}
\cite{ref10} Consider a sequence $\mathbf{z}$ of length $N=rm^2$, where $r$ is any positive integer and $m$ is a square-free integer. For any integer $n \in [0,N-1]$, by denoting $\lfloor n/m \rfloor $ as $\beta$ ($\beta \in \mathbb{Z}_{rm}$) and $n - \beta m $ as $\gamma$ ($\gamma \in \mathbb{Z}_{m}$), the CAZAC sequence can be constructed by
	\begin{eqnarray}
		\label{gene}
		z[n]=\exp(\textsf{j}\frac{2\pi g(\beta,\gamma)}{rm}),
	\end{eqnarray}
 where
	\begin{eqnarray}
		g(\beta,\gamma)=mc_{\rm{r}} \phi(\gamma)\beta^2 + \varphi(\gamma)\beta + \psi(\gamma).
	\end{eqnarray}
$c_{\rm{r}}$ is whether $1$ ($r$ is odd) or $1/2$ ($r$ is even),
	$\phi(\gamma)$: $\mathbb{Z}_{m} \rightarrow \mathbb{Z}_{r}$ is arbitrary function with $\gcd(\phi(\gamma), r) = 1$,
	$\varphi(\gamma)$: $\mathbb{Z}_{m} \rightarrow \mathbb{Z}_{rm}$ satisfies that $\{ \varphi(\gamma) (\operatorname{mod} m): \gamma \in \mathbb{Z}_{m} \}$ is a permutation over $\mathbb{Z}_{m}$, and $\psi(\gamma)$ is arbitrary real-valued function. 
	
	\label{lem-1}
\end{lemma}
 
\subsubsection{Parameter Design for General CAZAC Sequences}
\label{auto}
For simplicity, here we consider $\phi(\gamma)=\phi$ with $\gcd(\phi, r) = 1$. 
The magnitude of the cross-correlation $\left \| r_{z}[n] \right \|$ between the transmitted CAZAC sequence and the echo signal with the integer round-trip delay $\tau$ and the normalized Doppler shift $v$ can be derived as 
\begin{align}
\left \| r_{z}[n] \right \|\! =\!
 &\left \| \!\sum_{\beta=0}^{rm-1} \!\sum_{\gamma=0}^{m-1} \!z[\beta m\!+\!\gamma \!- \! \tau]z^*[\beta m\!+\!\gamma \! - \! n]e^{\textsf{j}2\pi {v}(\beta m+\gamma\!)} \!\right \|  \nonumber \\
\!	=\! &\left \| \sum_{\gamma=0}^{m-1} \!\sum_{\beta=0}^{rm-1}\!\!z[\beta_{\tau} m\!+\!\gamma_{\tau}]z^*[\beta_{n} m\!+\!\gamma_{n}]e^{\textsf{j}2\pi {v}(\beta m+\gamma)} \right \|  \nonumber  \\ 
	\leqslant &  \sum_{\gamma=0}^{m-1} \left \| \frac{\sin\left(\pi x(\gamma,n) \right) }{\sin\left(\frac{\pi}{rm} x(\gamma, n) \right)} \right \|,  \label{auto-co}
\end{align}
where $\beta m+\gamma - \tau$ is denoted as $\beta_{\tau} m + \gamma_{\tau}$ with $\gamma_{\tau} \in \mathbb{Z}_{m}$, and $\beta m+\gamma \!- \! n$ is denoted as $\beta_{n} m + \gamma_{n} $ with $\gamma_{n} \in \mathbb{Z}_{m}$. 
Besides, $x(\gamma,n)$ is defined as
\begin{equation}
	\begin{aligned}
		x(\gamma,n)\!=\!\langle 2c_{r} m\phi (\beta_{n} \!-\! \beta_{\tau}) \! + \! \varphi(\! \gamma_{n}\!) \! - \! \varphi(\! \gamma_{\tau}\!) \! + \! vrm^2  \rangle _{rm}. \label{x}
	\end{aligned}
\end{equation}
The main peak corresponds to the case $n=\tau$, i.e., $(\beta_n, \gamma_n)=(\beta_{\tau}, \gamma_{\tau})$, while the sidelobes appear when $(\beta_n, \gamma_n) \neq (\beta_{\tau}, \gamma_{\tau})$. 

%
%
%
%

\begin{algorithm} [t]
	\caption{Parameter Optimization Algorithm for (\ref{problem2})}
	\label{alg1}
	\begin{algorithmic}[1]
		\REQUIRE  Sequence parameters $r$ and $m$, sensing range $D_{\rm{r}}$, maximum Doppler shift $\bar{v}$ .
		
		\FOR{$ \phi = 1:r-1 $ $\land$ $\gcd(\phi, r)=1$} 
		\FOR{$ a=0:\lfloor \frac{r}{m} \rfloor $}
		\STATE Construct the CAZAC sequence $\mathbf{z}$ according to (\ref{gene}) based on $\phi$ and $\varphi(\gamma) =\langle a m \gamma+\gamma\rangle_{rm}$;
		\STATE Calculate the cross-correlation and record the PSLR value $\text{P}(\phi, a) $ ;
		\ENDFOR
		\ENDFOR
		\STATE $\hat{\phi},\hat{a}=\arg\max\limits_{\phi, a} {\text{P}(\phi, a)}$ ;
		\RETURN $\hat{\phi}$,$\hat{a}$
		
	\end{algorithmic}
	
\end{algorithm}

According to (\ref{auto-co}), $\left \| r_{z}[n] \right \|$ can be approximated by its upper bound, which is the sum of $ f(x) = \left\| \frac{\sin (\pi x(\gamma,n))}{\sin(\frac{\pi}{rm} x(\gamma,n))}\right \|$ for $\gamma = 0, 1,..., m-1$.
Inspired by the root index design of ZC sequences, the parameters $\phi$ and $\varphi(\cdot)$ are optimized to maximize PSLR within the scope of RoI under the maximum Doppler shift $\bar{v}$, which can be formulated as
\begin{equation}
	\begin{aligned}
		\label{problem2}
		&\max_{\phi, \varphi(\cdot)} \quad \!\frac{\left \| r_{z}[\tau]\right \| }{\max\limits_{ n \neq \tau} \left \| r_{z}[n]\right \|}, \text{ for $0 < n < \frac{2D_{\rm{r}}}{cT_s} $ }  \\
		\text{s.t.} \!\!\! \quad &\gcd(\phi, r)=1,\quad\! \!\!\{ \varphi(\gamma) (\operatorname{mod} m): \gamma \in \mathbb{Z}_{m} \}={\mathbb{Z}_{m}}.
	\end{aligned}
\end{equation} 

%

Exhaustive searching on the optimal $\phi$ and $\varphi(\cdot)$ may be impractical, which involves high computational complexity due to huge search space of $\varphi(\cdot)$.
To reduce the complexity, we consider a specific form of $\varphi(\cdot)$, which can be expressed as
\begin{align}
	\varphi(\gamma) =\langle a m \gamma+\gamma\rangle_{rm}, \quad
	a \in \{0, 1, \cdots, \lfloor r/m \rfloor\}.\label{form}
\end{align}
Since we have $\varphi(\gamma)\ \equiv \gamma(\operatorname{mod} m) $, $\{ \varphi(\gamma) (\operatorname{mod} m): \gamma \in \mathbb{Z}_{m} \}$ constitutes a permutation over $\mathbb{Z}_{m}$, satisfying the constraint in (\ref{problem2}).
By restricting the form of $\varphi(\cdot)$ to (\ref{form}), the searching on $\varphi(\cdot)$  can be transformed to the searching on $a$, which reduces the computational complexity from $\mathcal{O}(m!r^m)$ to $\mathcal{O}(r)$.

Following this philosophy, a low-complexity searching algorithm is proposed for the joint optimization of the parameters $\phi$ and $a$, which is shown in Algorithm~\ref{alg1}.
Firstly, the CAZAC sequences based on different values of $\phi$ and $a$ are constructed, respectively. After that, the cross-correlation results between each sequence realization and its echo counterpart under the maximum Doppler shift $\bar{v}$ are calculated. The PSLR values corresponding to different sequences within the scope of RoI are recorded, which is denoted as $\text{P}(\phi, a)$. Finally, the values of $\phi$ and $a$ attaining the maximal $\text{P}(\phi, a)$ are adopted for choosing the radar sequence.

\vspace{-2mm}
\section{Numerical Results}
\begin{figure}[tp!] 
\vspace{-5mm}	
 
\begin{center}
\includegraphics[width=.48\textwidth]{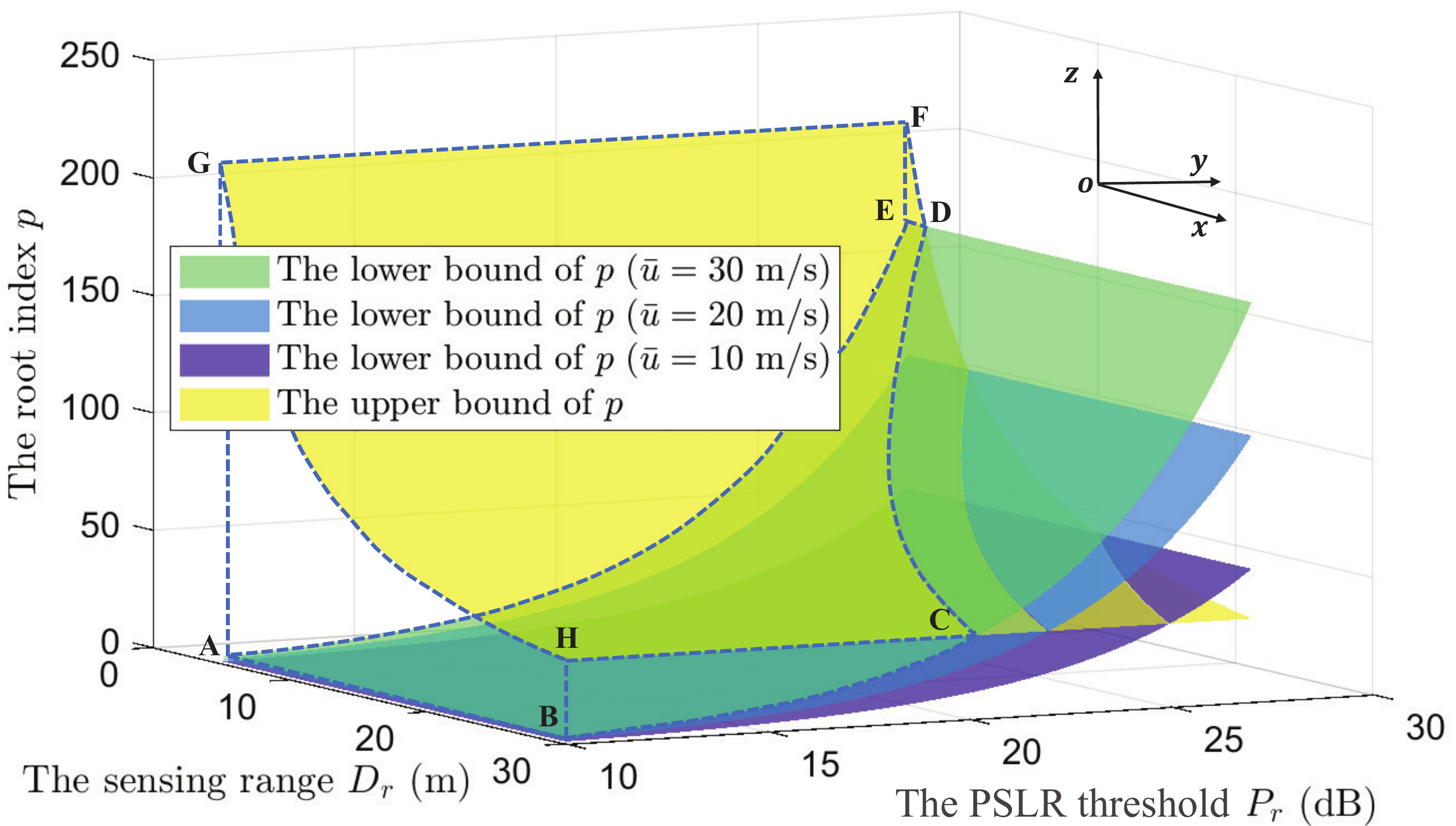}
\end{center}
\vspace{-5mm}	
\caption{The proposed feasible ranges of $p$ with respect to $P_r$ and $D_{\rm{r}}$, 
under different settings of maximum Doppler shifts.}
\label{sim1_1}
\vspace{-4mm}	
 
\end{figure}
In this section, a sub-mmWave sensing system with the carrier frequency of $240$ GHz and the sampling period of $0.2$ ns is considered. Firstly, we evaluate the feasibility and superiority of our proposed parameter design methodology for both ZC sequences (the proposed root index design) and general CAZAC sequences (the proposed $(\phi,a)$ design). After that, for multi-target sensing scenarios, the sensing performance of ZC sequences with the proposed root index design is evaluated in terms of the ROC curve, compared with its existing Doppler-resilient counterpart\cite{ref11}.

Figure~\ref{sim1_1} illustrates the proposed feasible ranges of the root index $p$ with respect to the sensing range $D_{\rm{r}}$ and the PSLR threshold $P_r$ for ZC sequences ($N=35537$) under different speed limits. 
The yellow surface, according to (\ref{solution}), is the upper bound of the feasible ranges, which is independent of the speed limit, whilst the green, blue and purple surfaces correspond to the lower bounds of the feasible ranges under different speed limits.
Taking $\bar{u}=30$ m/s as an example, the 3-dimensional volume ($\overline {\text{ABCDEFGH}}$) formed by the yellow and green surfaces is the feasible region of $p$, whose projection area in $xy$ plane corresponds to the feasible region of ($P_r$, $D_{\rm{r}}$), i.e., the set of ($P_r$, $D_{\rm{r}}$) realizations which ensure at least one solution of the root index.
As expected, with the increase in the requirement for sensing capability, i.e., a higher PSLR within a longer sensing range, the feasible range of $p$ narrows or even becomes an empty set.
Despite this, our proposed root index design is capable of finding an appropriate root index to achieve superior PSLR in most short-range radar applications even under severe Doppler shifts, e.g., over $20$ dB PSLR for $D_{\rm{r}}<30$ m under $\bar{u}=30$ m/s.

\begin{figure}[tp!] 
	\vspace{-3.5 mm}	
	
	\begin{center}
		\includegraphics[width=.49\textwidth]{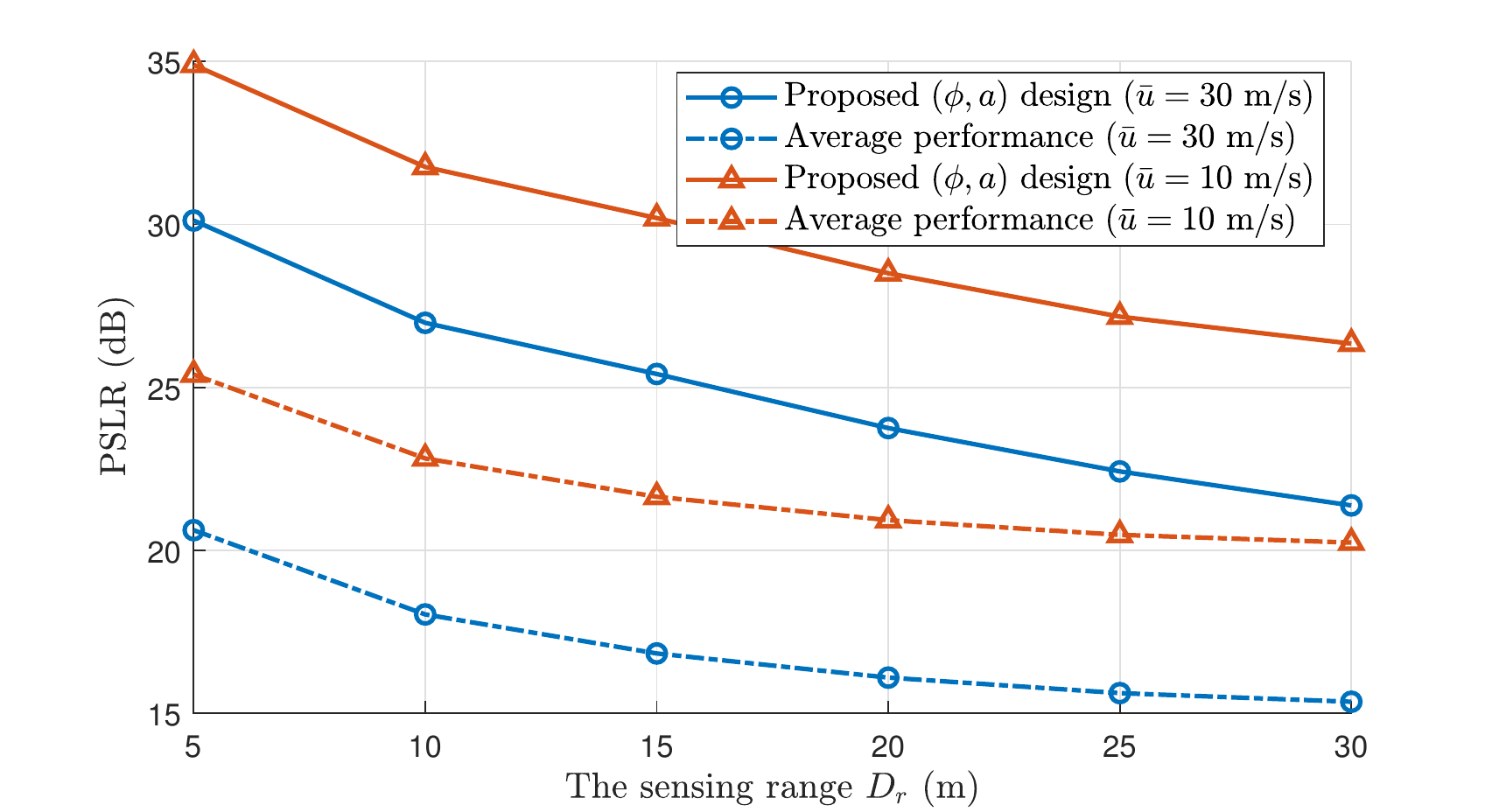}
	\end{center}
	\vspace{-5mm}	
	\caption{PSLR performance comparison between the proposed $(\phi,a)$ design and the average level for general CAZAC sequences.}
	\label{sim1_2}
	\vspace{-4mm}	
	
\end{figure}

For the general form of CAZAC sequences with $r=1009$ and $m=3$, the PSLR obtained by our proposed parameter $(\phi,a)$ design is investigated under different speed limits, as shown in Fig.~\ref{sim1_2}. The average PSLR performance of different parameter settings {among the whole search scope} is considered as the benchmark. {Specifically, the parameters $\phi$ and $[\varphi(0), \varphi(1), \varphi(2)]$ are randomly selected under the constraint conditions as (\ref{problem2}). The average performance is derived as the average PSLR of sequences corresponding to $10^4$ different sets of parameters.}
As sensing range increases, it becomes gradually difficult to maintain high PSLR within RoI. It is apparent that the PSLR of our proposed $(\phi,a)$ design is improved by at least $7$ dB under different Doppler shifts, compared with the average performance, which validates the superiority of our proposed parameter optimization algorithm.

\begin{figure}[tp!] 

\begin{center}
\includegraphics[width=.49\textwidth]{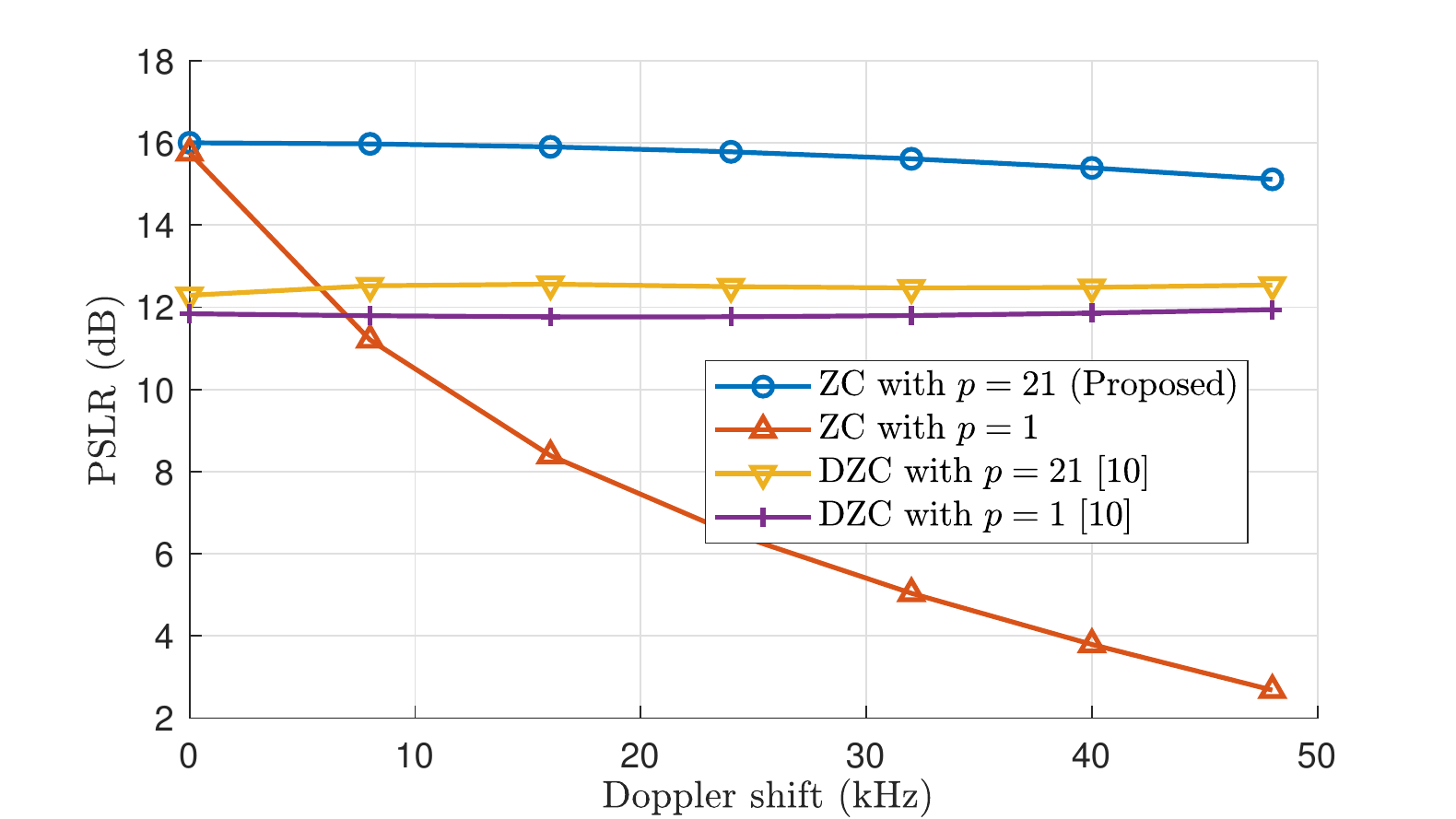}
\end{center}
\vspace{-5mm}	
\caption{The PSLR comparison of ZC sequences with/without the proposed root index design and the DZC sequences.}\label{simpslr}

\vspace{-4mm}	
 
\end{figure}
{The PSLR comparison of different sequences with respect to the Doppler shift is illustrated in Fig.~\ref{simpslr}. The SNR and sensing range are set as $-5$ dB and $50$ m, respectively. By substituting $D_{\rm{r}}=50$ m into (\ref{solution1}), $p=21$ is selected by our proposed root index design.
It can be seen that the PSLR of the proposed ZC sequences and the DZC sequences are more resilient to the Doppler shift than the ZC sequences with $p=1$. Besides, since the PSLR degradation of DZC sequences is mainly caused by noise amplification, it can not be improved significantly by applying the root index design algorithm to the DZC.}

\begin{figure}[tp!] 
\vspace{-3mm}	

\begin{center}
\includegraphics[width=.49\textwidth]{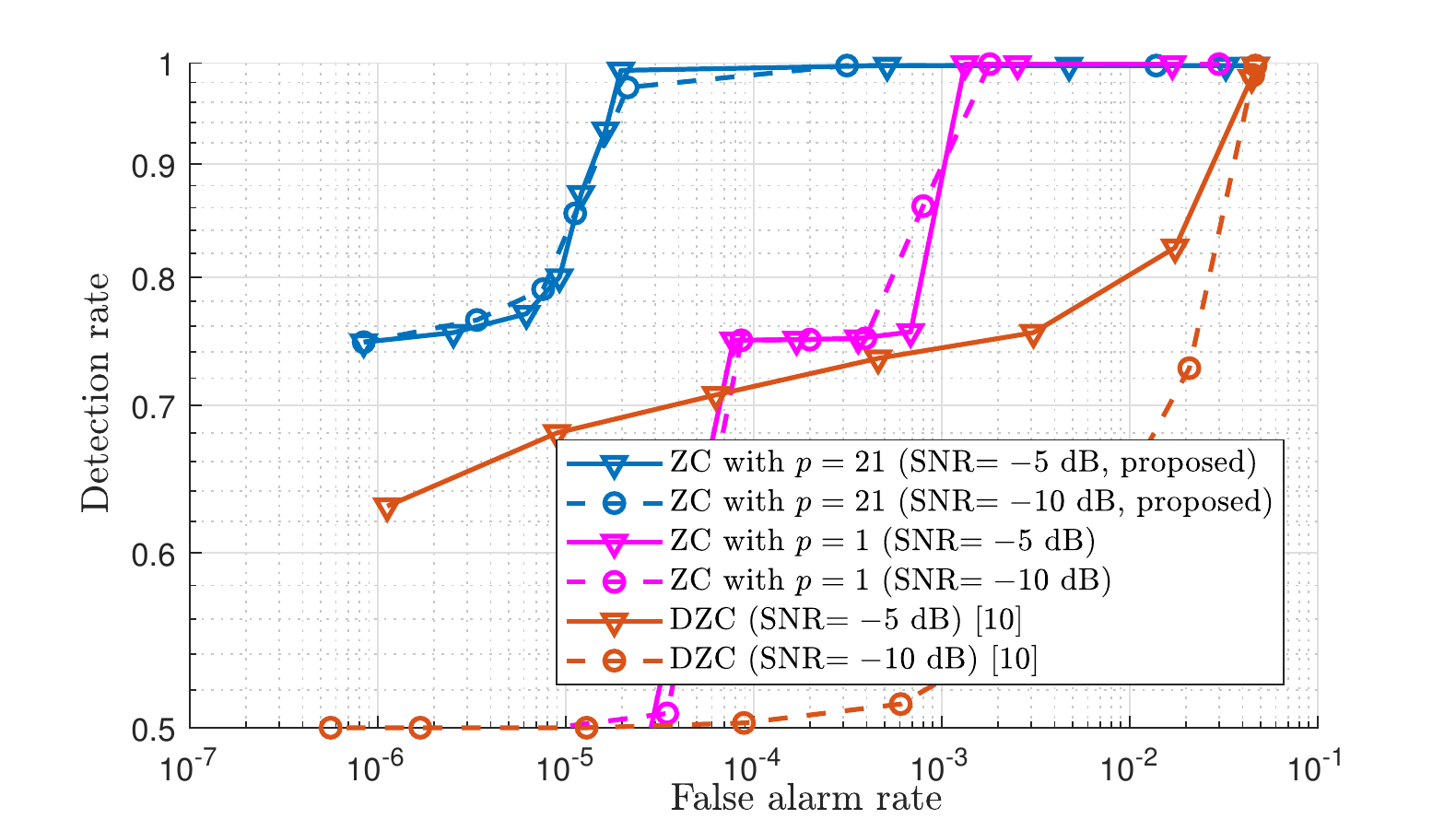}
\end{center}
\vspace{-5mm}	
\caption{The ROC curves of the proposed root index design and the DZC benchmark.}
\label{sim2}
\vspace{-4mm}	
 
\end{figure}

Besides, the ROC curves of {ZC sequences with/without the proposed root index design} and the DZC sequence are evaluated in Fig.~\ref{sim2}, where the receiving SNR is set as $-5$ or $-10$ dB.
The number of targets is set as $4$, and the relative distance and velocity to the source vehicle of each target are randomly distributed in $[0,50]$ m and $[-20,20]$ m/s, respectively. 
{Accordingly, the normalized Doppler shift ${v}$ is limited in $[-6.4\times 10^{-6},6.4\times 10^{-6}]$. By searching the Doppler shift in the specific range, the ambiguity problem can be avoided.}
Besides, the length of all sequences is set as $N=35537$ and they are assumed to be continuously transmitted for $K=100$ times. 
{The work in \cite{ref11} proposed two range estimation methods for DZC sequences, which are differential correlation and maximum likelihood joint estimation. For a fair comparison, the differential correlation method is employed in the simulations, which has the similar operational procedure and computational complexity to the circular correlation for ZC sequences.}
{Compared with the regular ZC sequence ($p=1$), the proposed root index design provides a preferable ROC curve because of the PSLR improvement.}
Besides, the ROC curve of the DZC is inferior to that of the proposed design. 
This is because, the DZC design utilizes differential encoding to improve the Doppler resilience of ZC sequences, which requires the decoding operation at the receiving end.
By applying the decoding, the noise power doubles, leading to degradation of the detection rate at low SNR. 
Nevertheless, our proposed root index design only adjusts the root index of the sequence without performing extra operations.
It maintains a high detection rate while reducing the false alarm rate, exhibiting better sensing performance than its DZC counterpart.

\begin{figure}[tp!] 
\vspace{-3mm}	

\begin{center}
\includegraphics[width=.49\textwidth]{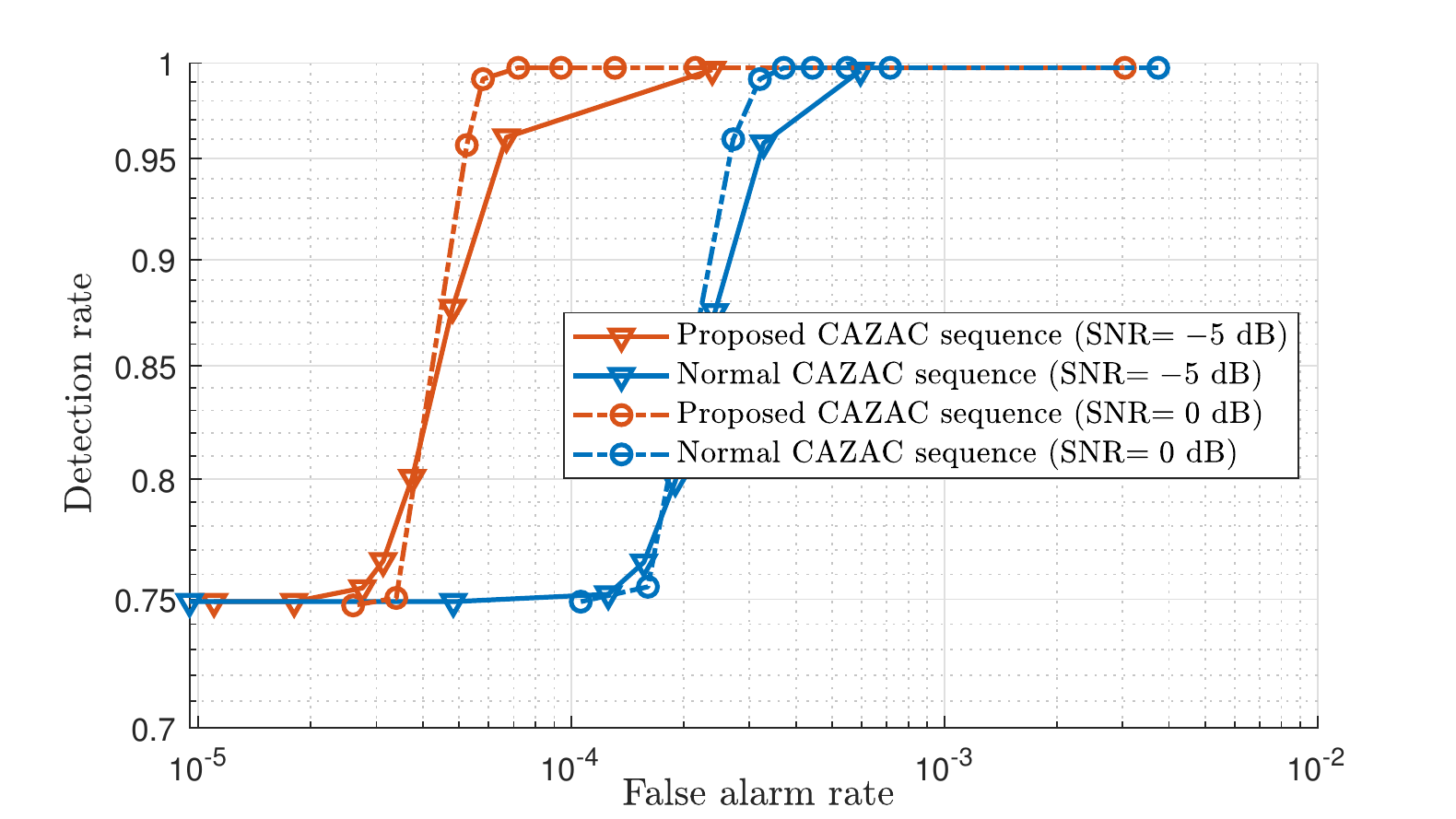}
\end{center}
\vspace{-5mm}	
\caption{The ROC curves of general CAZAC sequences with and without the proposed  $(\phi, a)$ design.}
\label{cazac_roc}
\vspace{-4mm}	
 
\end{figure}

{Finally, ROC curves of general CAZAC sequences ($r=1009,m=3$) with and without the proposed  $(\phi, a)$ design are compared in Fig.~\ref{cazac_roc}. Considering $D_r = 50$ m, the proposed  $(\phi, a)$ design ($\phi=181, a=120$) can be derived according to Algorithm 1. The normal CAZAC sequence is set to $\phi=181$ and $[\varphi(0), \varphi(1), \varphi(2)]= [421, 816, 276]$, whose PSLR is equal to the average PSLR of all sequences. It can be seen that at the same detection rate, the false alarm rate is reduced significantly by  the proposed $(\phi, a)$ design.}

\section{Conclusion}
At mmWave/THz frequencies, the sensing applications using CAZAC sequences are fragile to the severe Doppler shift effects, especially for high-mobility scenarios.
To solve this challenging issue, this paper proposes a parameter design methodology for CAZAC sequences to improve their resilience to Doppler shifts. 
Firstly, ZC sequences, as one of the most popular CAZAC sequences, are investigated. Based on number theory, the root index of the arbitrary-length ZC sequence is well-designed to satisfy the required PSLR threshold within the scope of RoI under Doppler shift effects.
After that, by extending the design philosophy to the general form of CAZAC sequences, a low-complexity searching algorithm is proposed to optimize the sequence parameters for PSLR improvement within the scope of RoI.
Finally, numerical results are provided to validate the feasibility and superiority of our proposed parameter design methodology.

 {Moreover, the proposed Doppler-resilient sequence design can be further improved in the following aspects. 
Firstly, 
to reduce the complexity, a quasi-optimal solution is derived by reducing the search scope in the proposed algorithm. To get the optimal solution, more advanced optimization algorithms (e.g., MM algorithm) require exploration.
Besides, the theoretical derivation in our paper is based on the assumption of $vN<1$, which fits well with many practical scenarios. However, 
the increasing estimation error and ambiguity under $vN > 1$ can be further investigated.}



%
%

\ifCLASSOPTIONcaptionsoff
  \newpage
\fi

%


\end{document}